\documentclass[preprint,superscriptaddress,aps]{revtex4}
\usepackage{amsfonts}
\usepackage{graphics}
\usepackage{graphicx}

\newcommand{\be}{\begin{equation}}
\newcommand{\ee}{\end{equation}}
\newcommand{\bea}{\begin{eqnarray}}
\newcommand{\eea}{\end{eqnarray}}
\newcommand{\pa}{\partial}

\begin{document}
\title{On the many-field $f(R)$ brane}

\author{D. Bazeia}
\email{bazeia@fisica.ufpb.br}
\affiliation{Instituto de F\'\i sica, Universidade de S\~ao Paulo\\
Caixa Postal 66318, 05315-970, S\~ao Paulo, SP, Brazil}
\affiliation{Departamento de F\'{\i}sica, Universidade Federal da Para\'{\i}ba\\
 Caixa Postal 5008, 58051-970, Jo\~ao Pessoa, Para\'{\i}ba, Brazil}

\author{R. Menezes}
\email{rmenezes@dce.ufpb.br}
\affiliation{Departamento de Ci\^{e}ncias Exatas, Universidade Federal da Para\'{\i}ba\\
58927-000, Rio Tinto, Para\'{\i}ba, Brazil}

\author{A. Yu. Petrov}
\email{petrov@fisica.ufpb.br}
\affiliation{Departamento de F\'{\i}sica, Universidade Federal da Para\'{\i}ba\\
 Caixa Postal 5008, 58051-970, Jo\~ao Pessoa, Para\'{\i}ba, Brazil}

\author{A. J. da Silva}
\email{ajsilva@fma.if.usp.br}
\affiliation{Instituto de F\'\i sica, Universidade de S\~ao Paulo\\
Caixa Postal 66318, 05315-970, S\~ao Paulo, SP, Brazil}


\begin{abstract}
We consider the Randall-Sundrum braneworld theory with a single extra dimension of infinite extent to investigate generalized $f(R)$ braneworld models in the presence of  several real scalar fields. In particular, we solve the modified Einstein equations for the case of flat brane, with zero cosmological constant, and for the case of bent brane, with a nonvanishing cosmological constant. In both cases we found explicit solutions for the scalar fields with analytical expressions for the respective warp factors. 
\end{abstract}

\maketitle 

The Randall-Sundrum (RS) theory \cite{RS} is one of the most interesting modern concepts in high energy physics. In the RS work \cite{RS}, the authors propose a braneworld model with a single extra dimension of infinite extent. Following this concept, a key idea in the cosmological description of the Universe is played by branes, so that the usual fundamental interactions can propagate only in the brane, the exception being gravity, which can propagate in the bulk space. As a result, the Universe is described by a 3-brane embedded into a higher-dimensional space, which is generally suggested to be five-dimensional, with the three usual spatial coordinates, one extra spatial coordinate and the time. The branes are known to solve the problems of the cosmological constant and the mass hierarchy \cite{RS}. 

In the RS theory, one describes a thin brane with one extra dimension of infinite extent, but we can also include scalar fields to make the brane thick \cite{gw,braneworld,bw1,melfo,bw2}. Several studies in this direction have been based on models including only one scalar field \cite{bw1,melfo,bw2}. Therefore, a natural development could consist, first, in the consideration of a modified gravity model (say, $f(R)$ model) instead of the usual Einstein gravity used in the original paper \cite{RS}, and second, in the introduction of a set of several scalar fields \cite{bentbrane}. 

Generalizations of the braneworld models have been carried out by several authors:  see, e.g., Refs.~\cite{frbrane,otherfrb,F,bentbrane,fof,O,review}. For instance, the application of a $f(R)$ modified gravity within the brane context has been studied for the first time in the paper \cite{frbrane} (and further, different aspects of such models have been considered in \cite{otherfrb}); the case of several scalar fields has been investigated in \cite{bentbrane}, where the first-order formalism, based on the reduction of the equations of motion to first-order differential equations (see also \cite{fof} for different examples of application of the first-order formalism within the gravity and cosmology contexts) has been successfully applied for solving the equations of motion, see also Ref.~\cite{O} for other interesting investigations on the subject.

A key issue concerning the first order formalism is the essential simplification of the equations of motion in the case of constant scalar curvature of the bulk, that is, if the bulk represents itself as either anti-de Sitter (adS), de Sitter (dS), or Minkowski space. Therefore, it would be interesting to consider models incorporating both these improvements, that is, modified gravity and the presence of the several dynamical scalar fields. From the physical point of view, this study would correspond to consider more generic cosmological models involving several types of matter within the modified gravity context, on the constant curvature background. This is just the problem we discuss in this work.

We start with the following action describing the $f(R)$ brane (cf. \cite{frbrane}):
\bea
S=\int d^4xdy\sqrt{-g}\left(-\frac{1}{4}f(R)+{\cal L}(\phi_1,\ldots,\phi_n)\right),
\eea
where $y$ is the extra coordinate. For the interested reader, we refer to \cite{review} for a recent, very interesting review on $f(R)$ theories. 
One can choose different forms for the function $f(R)$, with the restrictions that it should be continuous and differentiable and in the small curvature limit, reproduce the standard Einstein gravity with a cosmological term; also, it shouldn't have negative power in the scalar curvature, since in this case the model becomes unstable \cite{fof}. We could also consider even more generic actions involving other scalar invariants constructed from the Riemann tensor in the Lovelock type gravity \cite{Lovelock}, but as we will discuss below, this does not qualitatively change the general picture.

In the case of flat brane, the line element which controls the braneworld scenario is given by
\bea
\label{branemetric}
ds^2=e^{2A(y)}\eta_{ab}dx^adx^b-dy^2,
\eea 
  ${\cal L}$ is the Lagrange density restricted to represent scalar matter for simplicity. It has the form
\bea
{\cal L}=\frac{1}{2}g_{AB}\partial^A\phi^i\partial^B\phi^i-V(\phi_1,\ldots,\phi_n),
\eea
and it involves $n$ scalar fields (summation convention for repeated $i$-indices are assumed as well as for the space and time indices $A$ and $B$, with $A,B=0,1,...,4$), for the 5-dimensional metric tensor $g_{AB}$ defined by the line element above.  The scalar fields $\phi_i$ in these equations represent the simplest extension of the original braneworld model \cite{RS}; see, e.g., \cite{gw,braneworld,bentbrane}. We assumed that they only depend
on the fifth coordinate, but they contribute to modify the parameters of the standard model \cite{Terning}.
The modified Einstein equations and the scalar field equations look like
\bea
&&A^{\prime\prime}f_R-\frac{1}{3}A^{\prime}f^{\prime}_R+\frac{1}{3}f_R^{\prime\prime}=-\frac{2}{3}\phi^{\prime}_i\phi^{\prime}_i;\nonumber\\
&&(A^{\prime\prime}+A^{\prime 2})f_R-\frac{1}{8}f(R)-A^{\prime}f^{\prime}_R=-\frac{1}{4}\phi^{\prime}_i\phi^{\prime}_i+\frac{1}{2}V(\phi);\nonumber\\
&&\phi^{\prime\prime}_i+4A^{\prime}\phi^{\prime}_i=\frac{\pa V}{\pa\phi_i}.
\eea
The prime denotes derivative with respect to the extra dimension $y$, and $f_R\equiv {df(R)}/{dR}$. The scalar 
curvature is given in terms of the warp factor as
\bea
\label{scalcurv}
R=8A^{\prime\prime}+20(A^{\prime})^2. 
\eea
It is easy to check that for $f(R)=R$, the well-known results (see e.g. \cite{bentbrane}; similar equations have been discussed before in \cite{Gremm}, within the domain wall context) are reproduced.

Let us comment on the structure of the equations of motion in more generic Lovelock's \cite{Lovelock} gravitational models, involving other scalar contractions of the Riemann tensor. For the class of geometries that we considered,  it is straightforward to see that $R^2=
64(A^{\prime\prime})^2+320A^{\prime\prime}(A^{\prime})^2+400(A^{\prime})^4$, and that: $R_{ab}R^{ab}=
20(A^{\prime\prime})^2+64A^{\prime\prime}(A^{\prime})^2+32(A^{\prime})^4$ and $R_{abcd}R^{abcd}=4(A^{\prime\prime})^2+8A^{\prime\prime}(A^{\prime})^2+28(A^{\prime})^4$. Therefore, the linear combination of all invariants of the same order will have, qualitatively, the same structure. So, it is natural to expect that the results for models involving $f(R)$ and others, including functions of the other invariants, with the same power in the warp factor and its derivatives, will be similar up to numerical factors. We note however that this fact occurs for the very specific metric (\ref{branemetric}); in a more general geometry, the above similarity among the invariants $R^2$, $R_{ab}R^{ab}$ and $R_{abcd}R^{abcd}$ does not hold.

A first simplification that we can consider is the constant curvature case ($R'=0$ and so $f'_R=f''_R=0$) where the above equations become
\bea
&&A^{\prime\prime}f_R=-\frac{2}{3}\phi^{\prime}_i\phi^{\prime}_i;\\
&&A^{\prime 2}f_R-\frac{1}{8}f(R)=\frac{5}{12}\phi^{\prime}_i\phi^{\prime}_i+\frac{1}{2}V(\phi);\\
&&\phi^{\prime\prime}_i+4A^{\prime}\phi^{\prime}_i=\frac{\pa V}{\pa\phi_i}.
\eea
To get solutions of these equations we follow \cite{bentbrane}, writing the following first-order equation:
\bea
\label{sys1}
A^{\prime}(y)&=&W(\phi(y)).
\eea
By using this ansatz into Eq.~(6) we see that a possible solution can be cast to the form
\bea
\phi^{\prime}_i&=&-\frac{3}{2}f_R\frac{\partial W}{\partial \phi_i},
\eea
which substituted in Eq.~(7) gives the following restriction for the potential
\bea
\label{sys2}
V(\phi)&=&-\frac{15}{8}f^2_R\left(\frac{\partial W}{\partial \phi_i}\frac{\partial W}{\partial \phi_i}\right)+2f_R W^2-\frac{1}{4}f(R).
\eea
It is easy to verify that in the case of one scalar field, these equations reproduce the results of the paper \cite{frbrane}.

However, to solve these equations in a way consistent with the constant curvature condition, we must choose the potential $W(\phi_i)$ in a form compatible with the constant scalar curvature. Indeed, since the curvature is given by the Eq. (\ref{scalcurv}), and $A^{\prime\prime}=\frac{\partial W}{\partial\phi_i}\phi^{\prime}_i$, one can employ (\ref{sys1},\ref{sys2}) to relate the $W$ with the curvature through the equation
\bea
\label{eqpot}
\left(\frac{\partial W}{\partial \phi_i}\frac{\partial W}{\partial \phi_i}\right)-\frac{5}{3f_R}W^2+\frac{R}{12f_R}=0,
\eea
where the sum over repeated indices $i$ is assumed, and $R$ and $f_R$ are constants. For simplicity, let us consider the case $W=W(\Phi)$ with $\Phi=\alpha_i\phi_i=\alpha_1\phi_1+\alpha_2\phi_2+\ldots+\alpha_n\phi_n$. This dependence on $\phi_i$ suggests a symmetry among the several scalar fields and makes it possible to carry out an explicit and exact solution for the equations of motion. In this case, one has
\bea
a(W^{\prime})^2 +bW^2+c=0,
\eea
where $a=\alpha_i\alpha_i$, $b=-\frac{5}{3f_R}$, $c=\frac{R}{12f_R}$, and $W^{\prime}$ is the derivative of $W$ with respect to its complete argument $\Phi$.  It is clear that natural solutions of these equations are trigonometric, exponential and hyperbolic potentials, that is, either $W=B\sinh(\alpha_1\phi_1+\alpha_2\phi_2+\ldots+\alpha_n\phi_n)$, or $W=B\cosh(\alpha_1\phi_1+\alpha_2\phi_2+\ldots+\alpha_n\phi_n)$, or 
$W=B\sin(\alpha_1\phi_1+\alpha_2\phi_2+\ldots+\alpha_n\phi_n)$, or $W=B\cos(\alpha_1\phi_1+\alpha_2\phi_2+\ldots+\alpha_n\phi_n)$, or $W=B\exp(\alpha_1\phi_1+\alpha_2\phi_2+\ldots+\alpha_n\phi_n)$. Therefore, let us restrict ourselves, for example, to the case of two scalar fields, and test these possibilities.

First, we try the hyperbolic case: $W=B\sinh(\alpha\phi_1+\beta\phi_2)$ (in the case of only one field, this solution has been considered in \cite{frbrane}). In this case, the equations for the scalar fields are
\bea
\phi^{\prime}_1&=&-\frac{3}{2}f_R\alpha B\cosh (\alpha\phi_1+\beta\phi_2);\nonumber\\
\phi^{\prime}_2&=&-\frac{3}{2}f_R\beta B\cosh (\alpha\phi_1+\beta\phi_2)
\eea
Multiplying the first by $\alpha$, the second by $\beta$ and adding the two equations, we find
\bea
\arctan\sinh\Phi=-\frac{3}{2}(\alpha^2+\beta^2)f_R B(y-y_0),
\eea
where $y_0$ is an integration constant. So
\be
W=B\sinh\Phi=-B\tan[\frac{3}{2}(\alpha^2+\beta^2)f_R B(y-y_0)].
\ee
Then, since $A^{\prime}=W$, we can also obtain the warp factor
\bea
A^{\prime}(y)=-B\tan[C(y-y_0)],
\eea
where $C=\frac{3}{2}(\alpha^2+\beta^2)f_R B$.
Integrating this equation, one finds
\bea
A(y)=\frac{B}{C}\ln|\cos C(y-y_0)|=\frac{2}{3(\alpha^2+\beta^2)f_R}\ln\Biggl{|}\cos\left[\frac{3}{2}(\alpha^2+\beta^2)f_R B(y-y_0)\right]\Biggr{|}.
\eea
The potential $V(\Phi)$ becomes
\bea
V(\Phi)=\frac{5}{32}f_R R-\frac{1}{4} f(R)-\frac{9}{8}f_R W^2(\Phi).\nonumber\\
\eea

It is mandatory to verify that the above warp factor yields a constant scalar curvature; using the expression (\ref{scalcurv}), one finds
\bea
R=\frac{20B^2-8BC}{\cos^2 [C(y-y_0)]}-20B^2.
\eea 
The scalar curvature can be made constant by imposing the condition $5B=2C$. We note that the restriction to constant scalar curvatures is an important ingredient of the first-order formalism, allowing us to obtain a great number of solutions in explicit form \cite{fof}. Moreover, this choice rules out the singularity at $y=y_0$. The results is the negative constant $R=-20B^2$. This condition is satisfied if we have $\alpha^2+\beta^2={5}/{(3f_R)}$. It is clear that this solution can be straightforwardly generalized for the case of several scalar fields, with the differential equation for the $i$-th scalar field being $\phi^{\prime}_i=-\frac{3}{2}f_R\alpha_i B\cosh (\alpha_j\phi_j)$ (the sum over repeated indices is assumed), and then, everywhere in expressions for $\Phi$ and $A$, $\alpha^2+\beta^2$ being replaced by $\alpha_j\alpha_j$. We note that this potential is consistent with Eq.~(\ref{eqpot}) and with the condition of the constant (negative) scalar curvature. Thus, such a configuration is completely consistent.

Let us now try the trigonometric solution: $W=B\sin(\alpha\phi_1+\beta\phi_2)$. The equations for the scalar fields are
\bea
\phi^{\prime}_1&=&-\frac{3}{2}f_R\alpha B\cos (\alpha\phi_1+\beta\phi_2);\nonumber\\
\phi^{\prime}_2&=&-\frac{3}{2}f_R\beta B\cos (\alpha\phi_1+\beta\phi_2)
\eea
Introducing the field $\Phi=\alpha\phi_1+\beta\phi_2$ and integrating the equation, we arrive at
\bea
\Phi=\arcsin\tanh\left(\frac{3}{4}f_R(\alpha^2+\beta^2)B(y-y_0)\right),
\eea
which yields
\bea
W=B\tanh\left(\frac{3}{4}f_R(\alpha^2+\beta^2)B(y-y_0)\right).
\eea
Consequently, the warp factor is
\bea
A=\frac{B}{D}\ln\cosh (D(y-y_0))=\frac{4}{3f_R(\alpha^2+\beta^2)}\ln\cosh \left(\frac{3}{4}f_RB(\alpha^2+\beta^2)(y-y_0)\right),
\eea
with $D=\frac{3}{4}f_RB(\alpha^2+\beta^2)$. It is easy to check that the curvature in this case is also constant and negative, $R=-20B^2$, but in this case $2D=-5B$.
Again, the result can be straightforwardly generalized for the case of sum of an arbitrary number of the scalar fields.

In the above cases, if we replace $\sin\Phi$ by $\cos\Phi$, and $\sinh\Phi$ by $\cosh\Phi$, the curvature would stay constant, but positive, $R=20B^2$, and no brane world scenario would appear. Finally, for $W=B\exp\Phi$, the case of constant curvature yields $R=0$.

Up to now, we succeeded to apply the first-order formalism for the $f(R)$ modified gravity in the case of several scalar fields. For all these cases, we have found solutions, thus showing that the first-order formalism is a very powerful tool for the study of the RS braneworld model.

The above results can be generalized to the much harder case of non-zero cosmological constant ($\Lambda\neq0$), corresponding to bent branes \cite{F}. The metric in the case of the de Sitter space looks like \cite{Gremm}
\bea
ds^2=e^{2A(y)}\Biggl[dx^2_0-e^{2\sqrt{\Lambda}x_0}\sum\limits_{i=1}^3dx^2_i\Biggr]-dy^2,
\eea
and the scalar curvature now obeys
\bea
R=8A^{\prime\prime}+20(A^{\prime})^2-12\Lambda e^{-2A}.
\eea
The constant curvature condition yields the following solution for the warp factor:
\bea
\label{warp}
y=C_1+\int\frac{dA}{[C_2e^{-5A}+\Lambda e^{-2A}+\frac{R}{20}]^{1/2}},
\eea
where $C_1$ and $C_2$ are two constants (in principle one can choose $C_2=0$ but it will not essentially simplify the situation).

Then, for $R$ constant, the modified Einstein equations are reduced to
\bea
\label{sysmod}
f_R(R)\left[A^{\prime\prime}+4(A^{\prime})^2-3\Lambda e^{-2A}\right]-\frac{1}{2}f(R)=2\left(\frac{1}{2}\phi^{\prime}_a\phi^{\prime}_a+V(\phi)\right);\nonumber\\
f_R(R)\left[A^{\prime\prime}+(A^{\prime})^2\right]-\frac{1}{8}f(R)=-\frac{1}{2}\left(\frac{1}{2}\phi^{\prime}_a\phi^{\prime}_a-V(\phi)\right).
\eea
In these equations, one can eliminate $A^{\prime\prime}$ in favour of the curvature:
\bea
\label{sysmod1}
f_R(R)\left[\frac{R}{8}+\frac{3}{2}((A^{\prime})^2-\Lambda e^{-2A})\right]-\frac{1}{2}f(R)&=&\phi^{\prime}_a\phi^{\prime}_a+2V(\phi);\nonumber\\
4f_R(R)\left[\frac{R}{8}-\frac{3}{2}((A^{\prime})^2-\Lambda e^{-2A})\right]-\frac{1}{2}f(R)&=&-\phi^{\prime}_a\phi^{\prime}_a+2V(\phi).
\eea
Here we have multiplied the second equation by 4.
Now, since $A$ is an known (while implicit, see (\ref{warp})) function, one can find the solutions for $\phi_a$. We note that in the left-hand side of these equation, we have known functions, thus, one can find appropriate solutions for fields.

By taking into account that, for the warp factor (\ref{warp}), one has 
\be
\frac{3}{2}((A^{\prime})^2-\Lambda e^{-2A})=\frac{3R}{40}+\frac{3}{2}C_2e^{-5A},
\ee
 our system of equations is then reduced to
\bea
\label{sysmod2}
f_R(R)\left[\frac{R}{5}+\frac{3}{2}C_2e^{-5A}\right]-\frac{1}{2}f(R)&=&\phi^{\prime}_i\phi^{\prime}_i+2V(\phi);\nonumber\\
4f_R(R)\left[\frac{R}{20}-\frac{3}{2}C_2e^{-5A}\right]-\frac{1}{2}f(R)&=&-\phi^{\prime}_i\phi^{\prime}_i+2V(\phi).
\eea
The left-hand side of these equations are known functions; so, it remains to employ the equations of motion for the fields:
\bea
\phi^{\prime\prime}_i+4A^{\prime}\phi_i^{\prime}=\frac{dV}{d\phi_i},
\eea
where $A$ can be read off from (\ref{warp}). 
One can comment that for the set of $n$ scalar fields $\phi_i$, with $i=1\ldots n$, we have the system of $n+2$ equations, that is, $n$ equations for the scalar fields, and two equations from the system (\ref{sysmod2}). To provide the consistency of the system, we should solve it for $n+2$ variables, with $n$ of them being the scalar fields, one the curvature function $f(R)$, and the last one the potential $V(\phi)$; thus, different forms of $f(R)$ will correspond to different potentials.

There is also a special case $R=0$, corresponding to the warp factor satisfying the equation
\bea
A^{\prime}=\pm\sqrt{\Lambda} e^{-A},
\eea
that is, $A=\ln(\sqrt{\Lambda}(y-y_0))$, and $(A^{\prime})^2-\Lambda e^{-2A}=0$. In this case, one can try the function $f(R)=(a+bR)^n$, so, for the zero curvature, one has $f(R)\simeq a^n$, and $f_R(R)\simeq na^{n-1}$. The system (\ref{sysmod1}) is then reduced to
\bea
\label{sysmod3}
-\frac{1}{2}a^n&=&\phi^{\prime}_i\phi^{\prime}_i+2V(\phi);\nonumber\\
-\frac{1}{2}a^n&=&-\phi^{\prime}_i\phi^{\prime}_i+2V(\phi).
\eea
It is clear that the only solution in this case is the set of constant fields $\phi_i=const$. As we can see, for $R=0$ the results do not represent a braneworld solution.
The anti-de-Sitter case can be treated in a similar way, with no additional difficulty.

A natural continuation of the present study is to follow the lines of \cite{bentbrane}, attempting to solve the equations of motion numerically. Another study 
could correspond to the detailed consideration of the renormalization group flow, as investigated, for instance, in the second work in Ref.~\cite{bentbrane}. Also, in parallel to the conclusions of \cite{frbrane}, we hope that the $f(R)$ modification of gravity, in the case of several scalar fields coupled to it, would allow for a supersymmetric extension. We are planning to investigate these issues elsewhere. Several other studies on branes can be carried out, and we can, for instance, consider the $f(R)$ brane scenario studied in the present work, within the diversity of contexts explored in \cite{other}, including fermions and other fields. Also, we could use the present approach to generalize investigations \cite{me} which deal with interactions between the dark matter and dark energy sectors.

{\bf Acknowledgements.} We would like to thank CNPq and FAPESP for partial financial support. The work by A. Yu. P. is supported by the
CNPq project No. 303438/2012-6. R. M. thanks Instituto de F\'\i sica, Universidade de S\~ao Paulo, for the kind hospitality during a visit to conclude this work.

\end{document}